 \newlength\smallfigwidth
\documentclass[aps,prl,twocolumn,showpacs,superscriptaddress,groupedaddress]{revtex4-1}  

\usepackage{graphicx}  
\usepackage{dcolumn}   
\usepackage{bm}        
\usepackage{amssymb}   

\usepackage{graphicx,graphics}
\usepackage[tight,TABTOPCAP]{subfigure}
\usepackage{epstopdf}
\begin{document}

\preprint{UFES}

 \title{Topological Hall effect induced Skyrmion-Antiskyrmion coupling in inhomogeneous racetrack}

 \date{\today}

 \author{R. C. Silva} \email{rodrigo.c.silva@ufes.br}
\affiliation{Departamento de Ci\^encias Naturais, Universidade
Federal do Esp\'{i}rito Santo, S\~ao Mateus, 29932-540, Esp\'{i}rito
Santo, Brazil.}

 \author{R. L. Silva} \email{ricardo.l.silva@ufes.br}
\affiliation{Departamento de Ci\^encias Naturais, Universidade
Federal do Esp\'{i}rito Santo, S\~ao Mateus, 29932-540, Esp\'{i}rito
Santo, Brazil.}

 \author{A.R. Pereira} \email{apereira@ufv.br}
\affiliation{Departamento de F\'{i}sica, Universidade Federal de
Vi\c{c}osa, Vi\c{c}osa, 36570-000, Minas Gerais, Brazil.}

\begin{abstract}
In this paper we investigate a magnetic racetrack consisting of a junction of three materials with different properties. Indeed, this magnetic system is composed by two distinct regions (racetracks) connected by a thin interface: the first region (termed sector $1$) has isotropic in-plane magnetic chirality and supports skyrmion ($S$) excitations while the second (sector $3$) has anisotropic chirality and consequently supports antiskyrmions ($A$). The interface, which would be a third region (sector $2$, connecting sectors $1$ and $3$) located in the central part of the racetrack, is an easy-axis Heisenberg ferromagnetic material. The topological structures $S$ and $A$ are put in motion by applying a spin-polarized current. Under certain conditions, we show that the skyrmion and the antiskyrmion created in their respective sectors are simultaneously impelled to the interface (due to the Magnus force) to apparently become a unique object (a skyrmion-antiskyrmion pair or $SAP$). After glued by sector $2$, the skyrmion and the antiskyrmion move together (as a $SAP$) along the direction of the applied current. It is also shown that such an engineered racetrack can support a sequence of several $SAP$ structures in motion, forming a current.
\end{abstract}

\maketitle
\section{Introduction}
Over the last years magnetic skyrmions have attracted immense interest because of their scientific and technical appeal. These small spin textures have strong stability, characterized by a well defined topological charge that offers good conditions for skyrmions becoming information carriers in the field of spintronics/skyrmionics \cite{Bogdanov,Roszler,Yu,Sampaio,Fert}. These topological solitonic excitations may appear isolated or condensed in regular lattices and nowadays they can be directly visualized in chiral magnetic compounds \cite{Yu2010,Heise2011}. Such experimental observations have motivated studies on their creation, manipulation, and electric detection in the presence of electric currents as well as magnetic fields. Specifically, their structures come from the Dzyaloshinskii-Moriya interaction (DMI), which breaks the chiral symmetry of the magnetic object. The DMI results from the spin-orbit interaction and it is only non-zero for solids lacking bulk or structure inversion symmetry. Particularly, the skyrmions stabilized in systems with surface or interface-induced DMI seem to be more promising for applications in spintronics than bulk systems with DMI \cite{Hoffmann}. The interface offers a great variety of options for optimizing and controlling magnetic parameters: variation of the interface composition, interface crystal symmetry, the film thickness, as well as the fabrication of interlayers and multilayers \cite{Heide,Bergmann,Nandy, Dupe}.

\begin{figure}[hbt]
    \centering
    \subfigure[]{\includegraphics[width=40.0mm]{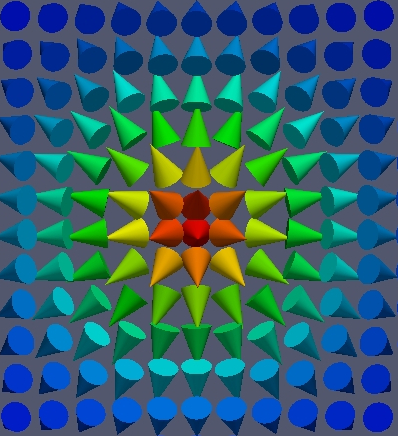}}
    \subfigure[]{\includegraphics[width=41.7mm]{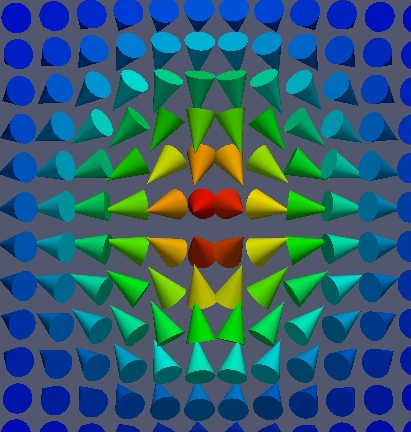}}
    \caption{(Color online) Top view of N\'{e}el-type skyrmion and antiskyrmion. The directions of the cones represent the magnetization and their colors, ranging from red ($m_{z}=+1$) to blue ($m_{z}=-1$), indicate the behavior of the out-of-plane spin component. (a) In a N\'{e}el-type skyrmion, the spins rotate in the radial planes from the core in which they are up to the periphery where they become down. (b) Antiskyrmions are instead of achiral spin textures without skyrmionic properties.}
    \label{sky1}
\end{figure}

\begin{figure}[hbt]
    \centering
    \includegraphics[width=80.0mm]{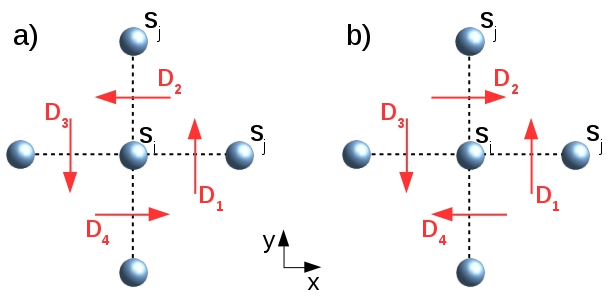}
    \caption {(a) Magnetic materials with DMI vectors on four $j$ sites with the same rotational sense stabilize skyrmion structures. (b) Magnetic materials with broken in-plane rotation symmetry yield anisotropic DMI which may stabilize antiskyrmions. }
    \label{any}
\end{figure}

The N\'{e}el-type hedgehog skyrmion, as shown in Fig. \ref{sky1}(a), originates from the interfacial DMI. Additionally, it may exist another distinct type of spin texture, an antiskyrmion (see Fig. \ref{sky1}(b)). Magnetic antiskyrmions are topologically nontrivial achiral spin quasiparticles that may occur in cases where the magnetic chirality is anisotropic, as opposed to skyrmions with isotropic in-plane chirality \cite{Huang,Nagaosa}. This kind of texture has been reported in certain tetragonal material with acentric crystal structure and $D_{2d}$ symmetry, in engineered $Co/Pt$ multilayers, and  in the $Mn-Pt-Sn$ inverse Heusler compound above room temperature \cite{Nayak}. It has also been predicted that interfacial DMI with $C_{2\nu}$ symmetry can lead to the formation of antiskyrmions in ultrathin magnetic films. For instance, it has recently been realized in epitaxial $Au/Co/W$ magnetic films \cite{Camosi}. Similar to the usual skyrmions, antiskyrmions may also induce a Magnus-force associated propagation derivation, i.e, an antiskyrmion Hall effect \cite{Sen}. The antiskyrmion Hall angle, between propagation direction and current direction, strongly depends on the applied current direction concerning the internal spin texture of antiskyrmions. The stabilization of skyrmions and antiskyrmions at the interface between a thin magnetic film and a heavy metal adjacent layer occurs due to the Dzyaloshinskii-Moriya vector $\vec{D}_{ij}$ between spins $\vec{S}_{i}$ and $\vec{S}_{j}$ on atomic sites $i$ and $j$ that lies within the film plane. For skyrmions stabilization, it is usually assumed that the interfacial DMI vectors on four $j$ sites have the same rotational sense (Fig. \ref{any}a, for isotropic DMI). In contrast, atomic configurations with broken in-plane rotation symmetry at the interface (Fig. \ref{any}b) may lead to anisotropic interfacial DMI, which stabilizes antiskyrmions.

\begin{figure}[hbt]
    \centering
    \includegraphics[width=9.0cm]{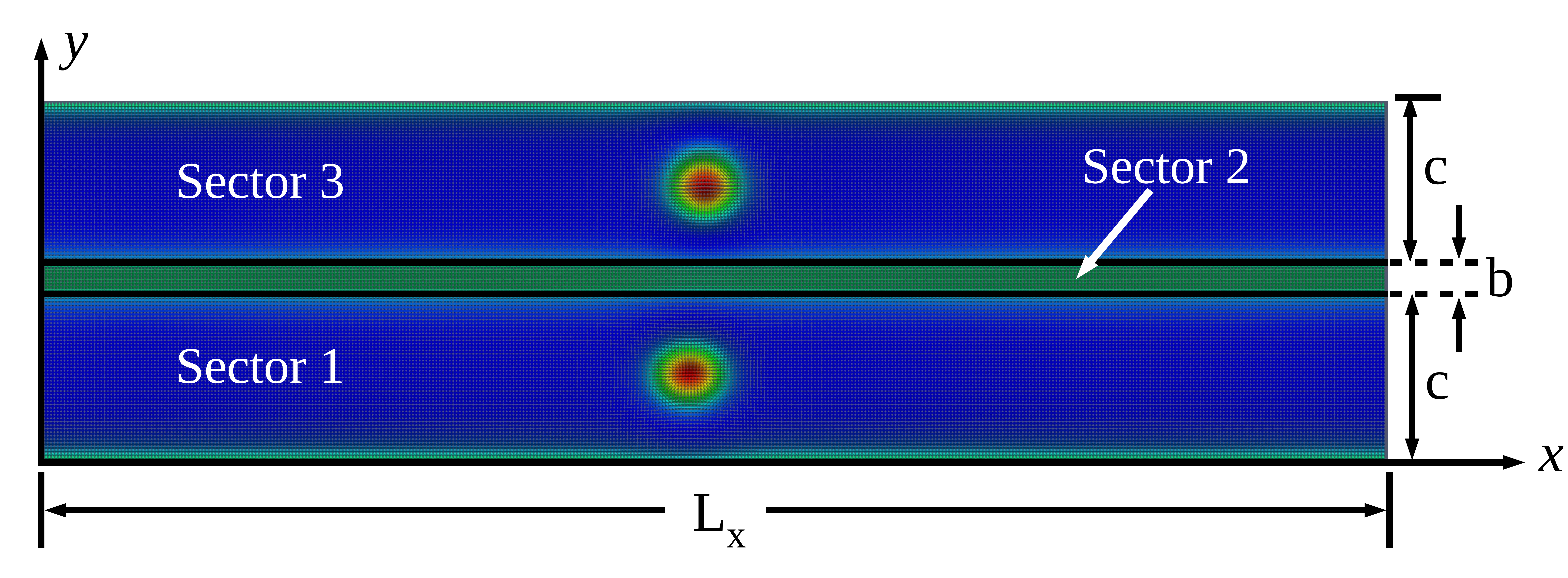}
    \caption{(Color online) An inhomogeneous racetrack engineered with $3$ sectors containing initially a skyrmion at sector $1$, an antiskyrmion at sector $3$. These two sectors are connected by a thin interface composed by a two-dimensional easy-axis ferromagnet (sector $2$) with width $b=6a$ ($a$ is the lattice spacing).}
    \label{racetrack}
\end{figure}

In this work, we focus on the possibility of composing these two objects, i.e., we examine a prospect able to form a stable  skyrmion-antiskyrmion pair. Therefore, since these topological objects cannot live in the same compound, we have to join two different materials to perform this task. Another important detail for achieving an attached skyrmion-antiskyrmion multifaceted structure is the fact that it can only exist in constant movement. So a permanent spin-polarized current has to be applied along the racetrack. When the current is turned off, the pair is dismembered in typical skyrmion and antiskyrmion excitations. With the advances in the field of nanotechnologies, it would be perfectly possible, in a tailor-designed system, to combine both types of materials of Fig. \ref{any}, joining them by a third magnetic compound, forming the desired inhomogeneous racetrack. Next, by exploiting the physical phenomena usually observed for these topological objects (movement induced by applied currents, topological Hall effect etc), we design a state of affairs that allows to stabilize jointly a skyrmion and an antiskyrmion which materialize in a novel spin texture (a skyrmion-antiskyrmion pair or $SAP$ in short). It is also shown that a current containing several $SAP$ textures is also possible in this engineered racetrack material.

\begin{figure*}
    \centering
    \subfigure[]{\includegraphics[width=8.5cm]{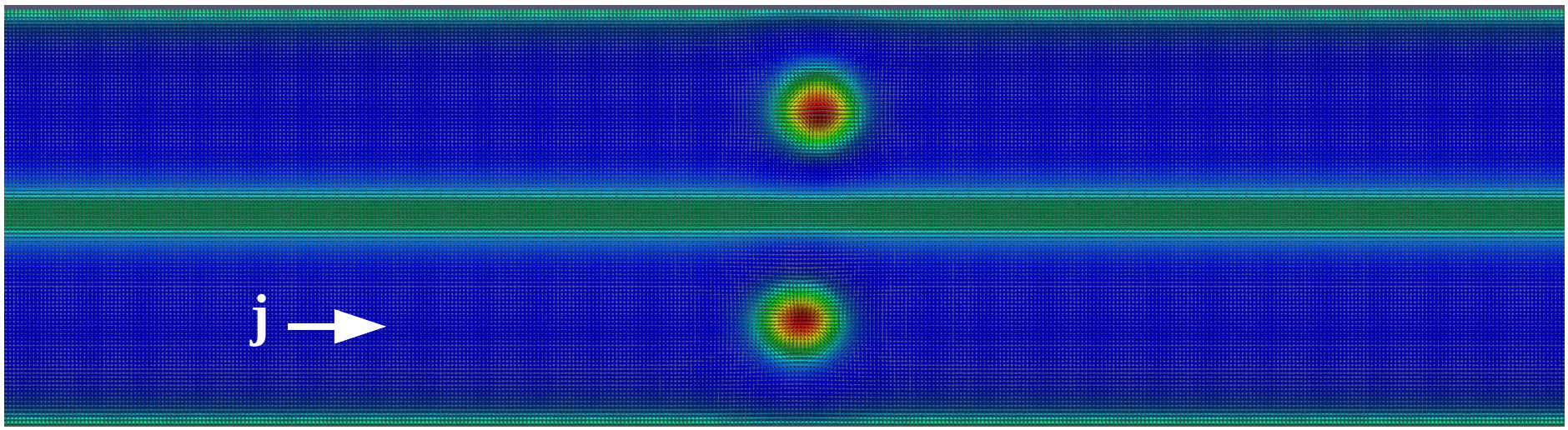}}
    \quad
    \subfigure[]{\includegraphics[width=8.5cm]{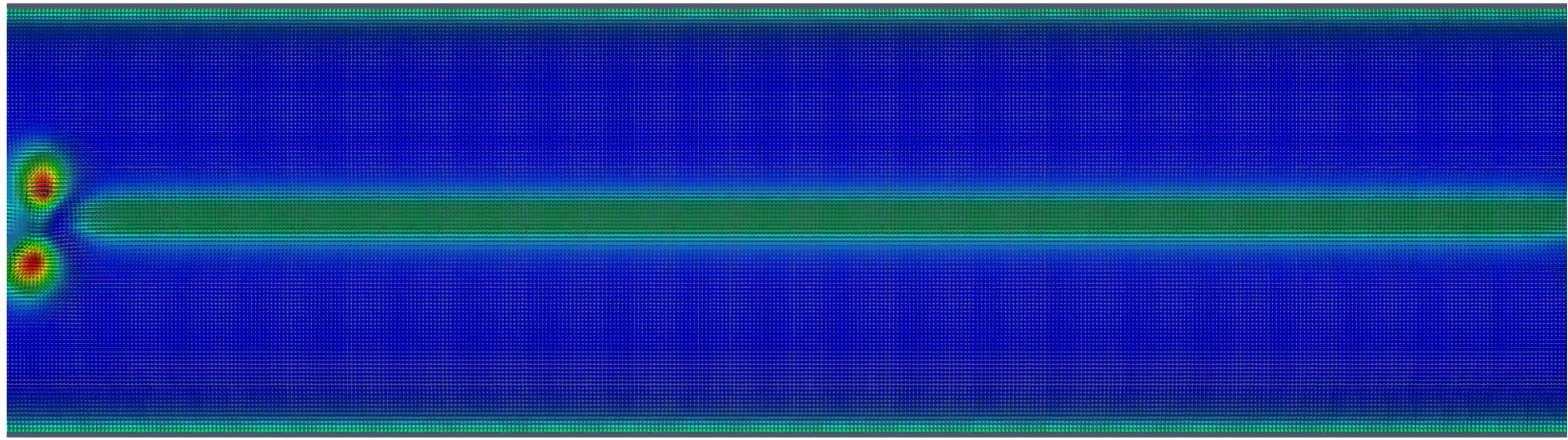}}
    \subfigure[]{\includegraphics[width=8.5cm]{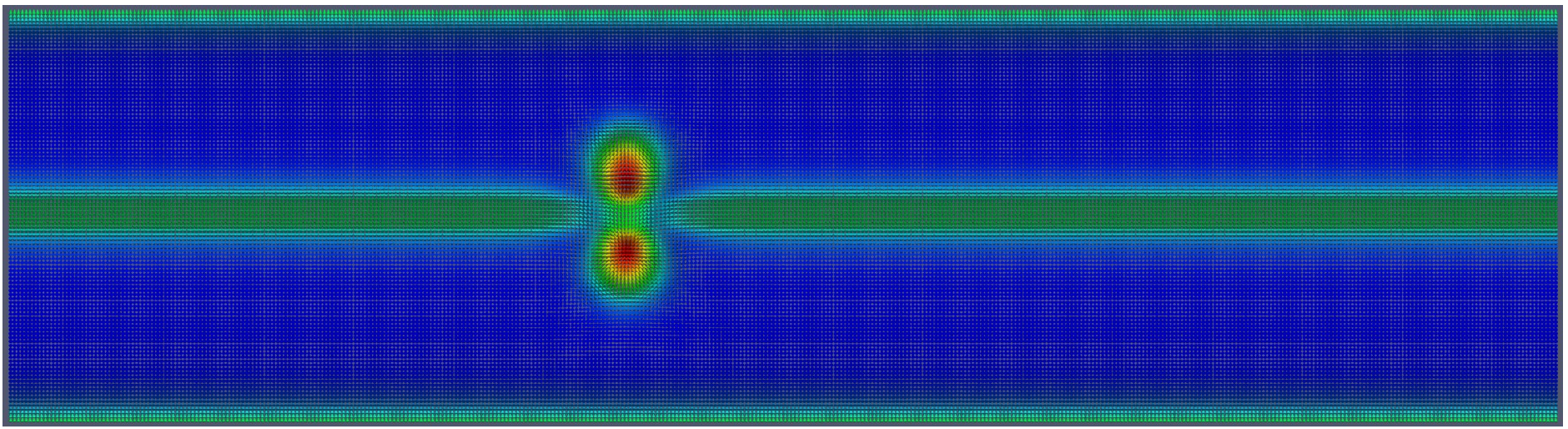}}
     \quad
    \subfigure[]{\includegraphics[width=4.2cm]{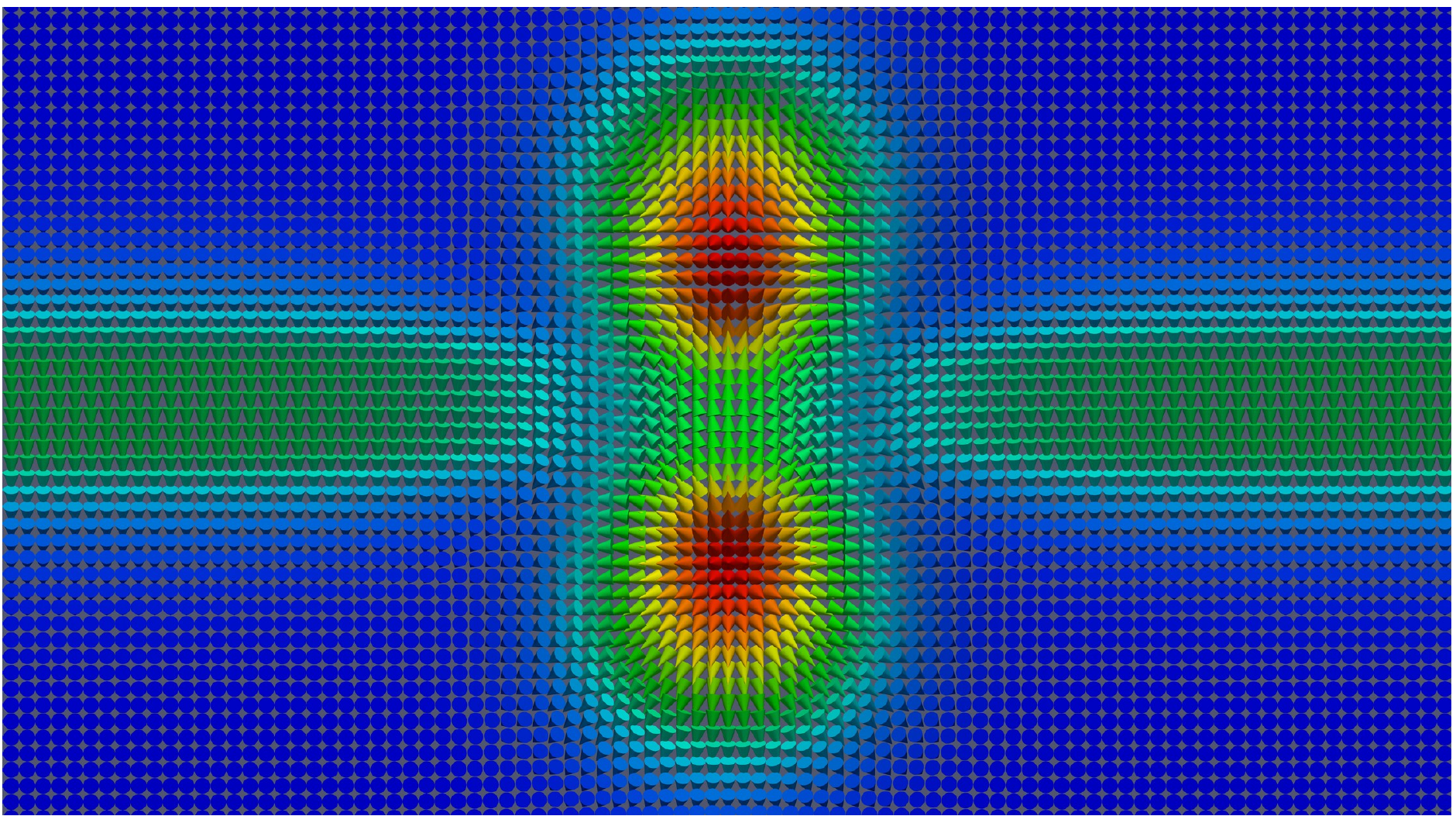}}
    \caption{(Color online) a) Initial configuration of a skyrmion and an antiskyrmion created at sectors $1$ and $3$ respectively. A spin current $\vec{\jmath}=0.10 \hat{x}$ is applied in order to put the topological objects in motion. b) Skyrmion and antiskyrmion are deflected to the interface (sector $2$) due to the Magnus force. c) Skyrmion and antiskyrmion form a pair which moves coherently along the interface. d) Zoom of the structure of a skyrmion-antiskyrmion pair ($SAP$) in the inhomogeneous racetrack considered here.}
    \label{current1}
\end{figure*}

\section{Model and Methods}
In order to describe a thin magnetic film we consider a two-dimensional lattice with magnetization vectors distributed in a regular rectangular lattice divided into three sectors as shown in Fig. \ref{racetrack}: (1) The skyrmion sector (or sector $1$), (2) Anisotropic ferromagnetic sector (or sector $2$), and (3) The antiskyrmion sector (or sector $3$). Sectors $1$ and $3$ are connected by the interface (sector $2$). In our model, the skyrmion and antiskyrmion energies are determined by the competition between the Heisenberg, Dzyaloshinskii-Moriya and Zeeman interactions together with the magnetic anisotropy energy, expressed in terms of the spin-lattice model applied to an interface geometry.
\begin{equation}
    \begin{array}{l}
        \mathcal{H} = -J\sum_{\langle i,j \rangle} \left(\vec{S}_{i}\cdot\vec{S}_{j}\right) - \sum_{\langle i,j \rangle}\vec{D}_{ij}\cdot \left(\vec{S}_{i}\times\vec{S}_{j} \right) - \vec{B} \cdot \sum_{i} \vec{S}_{i}\\
        -\sum_{i} K^{y}_{i}\left( \vec{S}_{i}\cdot\hat{e}_{y} \right)^{2}
    \end{array}\label{eq1}
\end{equation}
where the sum $\langle i,j \rangle$ is over nearest-neighbor spins. The classical spins $\vec{S}$ of length $S=1$ at atomic sites $(i, j)$ interact by the couplings $J$, $\vec{D}_{ij}$, and  $K^{y}_{i}$ is on-site interaction. The DMI term can be written as,
\begin{equation}
    \begin{array}{l}
        H_{DMI} = D_{x} \left[S_{ij}^{z} \left(S_{i+1,j}^{x}-S_{i-1,j}^{x}\right)-S_{ij}^{x}\left(S_{i+1,j}^{z}-S_{i-1,j}^{z}\right)\right] \\
        +D_{y} \left[ S_{ij}^{z} \left( S_{i,j+1}^{y}-S_{i,j-1}^{y} \right) -S_{ij}^{y}\left(S_{i,j+1}^{z}-S_{i,j-1}^{z}\right)\right],
     \end{array}
\end{equation}
where  $D_{x}=D_{y}$ at the sector $1$,  $D_{x}=D_{y}=0$ at the sector $2$, and $D_{x}=-D_{y}$ at the sector $3$. The first case is the isotropic DMI, which leads to skyrmion stabilization. The last case is the anisotropic DMI that favors opposite chirality along the $x$ and $y$ directions, which allows the stabilization of magnetic antiskyrmions. $K^{y}_{i}$ is the easy axis constant, nonzero only at the sector $2$. The magnetic field, $\vec{B}=-B_{z}\,\hat{z}$, is applied perpendicularly to all the sample with the same intensity.

In our simulations we take into account the ground state obtained from Eq.(\ref{eq1}). This is done by applying two different paths: first, we can work with a single pair of skyrmion and antiskyrmion. This configuration is obtained by spin dynamics technique, that relax from an initial randomly lattice. Second, several excitations (skyrmions or antiskyrmions) can be studied by a thermodynamic approach. For this, we have used a simulated annealing process, which is a Monte Carlo scheme where the temperature is slightly reduced in each step of the process in order to drive the system to a thermal equilibrium configuration. Our Monte Carlo procedure consists of a Hinzke-Nowak algorithm \cite{Evans}. We decrease the temperature from $T=1.0$ $J/k_{B}$ until $T=0.005$ $J/k_{B}$ with $\delta T = 0.005$ $J/k_{B}$ ($k_{B}$ is the Boltzmann constant). After that, we used the spin dynamics to relax the system, driving it to the zeroth temperature configuration.

The dynamics of the system was obtained by solving numerically the Landau-Lifshitz-Gilbert \cite{Landau, Gilbert} equation, given by
    \begin{equation} \label{LLGS}
        \begin{array}{l}
            \frac{d \vec{S}_{i}}{dt}=-\gamma \vec{S}_{i} \times \hat{H}_{eff}^{i} - \alpha \left(\vec{S}_{i} \times \frac{d \vec{S}_{i}}{dt}\right),
        \end{array}
    \end{equation}
where $\gamma$ is the gyromagnetic ratio, $\hat{H}_{eff}^{i}=-\frac{1}{S_{s}}\frac{\partial \mathcal{H}}{\partial \vec{S}_{i}}$ is the local effective field and $\alpha$ is the Gilbert damping constant. Here, we set $\gamma=1$ and $\alpha=0.1$. Fourth-order Runge-Kutta method has been used to solve Eq.\ref{LLGS}.

\begin{figure}[hbt]
    \centering
    \includegraphics[width=9.0cm]{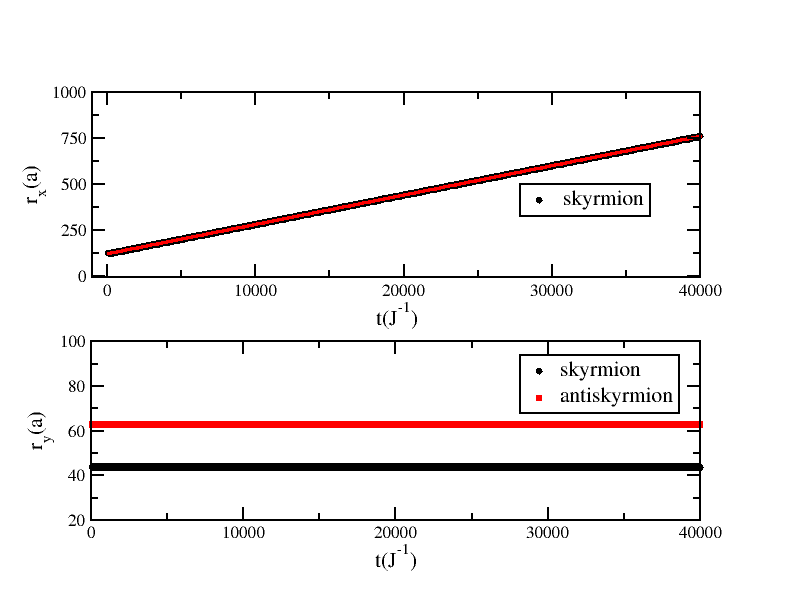}
    \caption{(Color online) Top: Horizontal position of the skyrmion ($S$) element of a $SAP$ as a function of time in a racetrack when the $SAP$ moves on the influence of a spin current. The red straight line represents a linear fit made on the $r_{x}$ data. The other integrant of a $SAP$ (antiskyrmion, $A$) follows the same line (not shown). Bottom: vertical positions of the $SAP$ elements: skyrmion (black circles) and antiskyrmion (red squares). These results show that the $SAP$ integrants $S$ and $A$ move together as coupled structures. They centers can be substituted by the mass center of the rigid $SAP$.}
    \label{position}
\end{figure}

\begin{figure}[hbt]
    \centering
    \includegraphics[width=9.0cm]{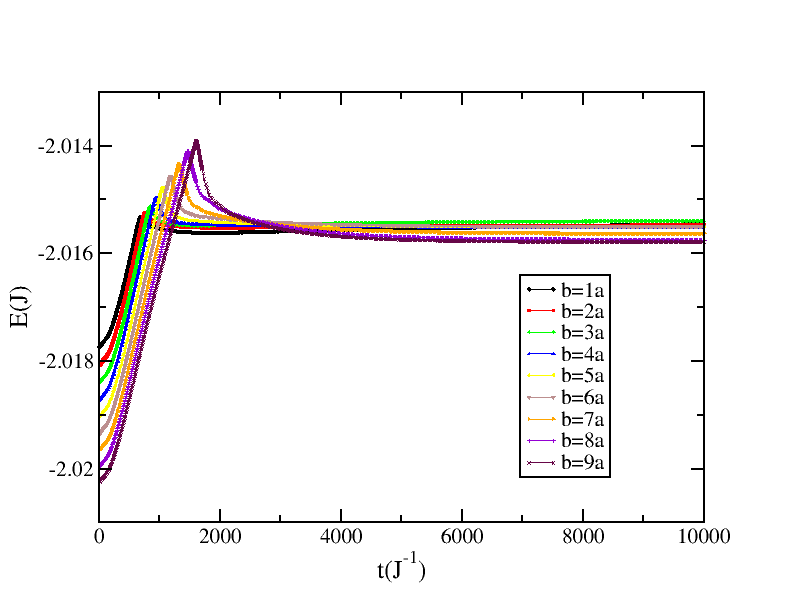}
    \caption{(Color online) Energy per spin of a racetrack with different interface (section $2$) widths as a function of time $t$.}
    \label{energy}
\end{figure}

After stabilizing the system, a spin current is applied giving rise to the Berger spin-transfer torque \cite{Brattas}:
    \begin{equation}\label{eq3}
        \vec{\tau}_{B}= p\left(\vec{\jmath}\cdot\nabla\right)\vec{S}\,,
    \end{equation}
    \begin{equation}\label{eq4}
        \vec{\tau}_{B\beta}= p\beta\vec{S}\times\left(\vec{\jmath}\cdot\nabla\right)\vec{S}\,,
    \end{equation}
where Eqs. (\ref{eq3}) and (\ref{eq4}) are the adiabatic and non-adiabatic torque, respectively. Here, $p$ is the spin polarization of the electric current density $\vec{\jmath}$, while $\beta$-parameter characterizes its relative strength to the Berger's torque ( Eq.(\ref{eq3})).  Our calculations have considered a racetrack with periodic boundary conditions (PBC) along \textit{x}-direction and open boundary condition (OBC) along \textit{y}-direction; we have also taken $p=-1$ and $\beta=0$. The choice of $\beta=0$ is due to the fact that the contribution of the non-adiabatic torque can be considered negligible as compared to its adiabatic counterpart\cite{Nagaosa2}.

\begin{figure}[hbt]
    \centering
    \includegraphics[width=9.0cm]{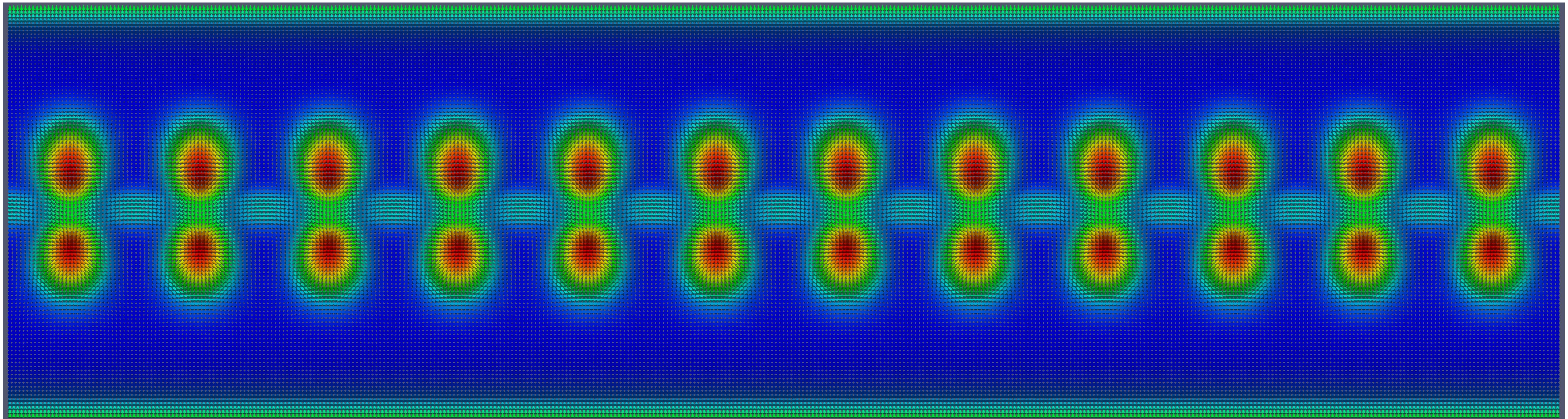}
    \caption{(Color online) A current of $SAP$ spin textures along the racetrack. This current propagates with constant velocity along the system. Here, the width of section $2$ is $b=6a$.}
    \label{current}
\end{figure}

\section{Results and Conclusions}
To develop our investigation, we take $J=1$ along all over the system, $D_{x}=D_{y}=0.4\, J$ at the sector $1$ and $D_{x}=0.4 \, J$ and $D_{y}=-0.4 \, J$ at the sector $3$. At the band of lattice that separates the skyrmion and antiskyrmion sectors (sector 2), we take $D_{x}=D_{y}=0$ and $K^{y}_{i}=0.05 \, J$ in its interior. At the upper and lower edges of the sector $2$, we have chosen an intermediary value to DMI, $D_{x}=D_{y}=0.2\, J$ in absolute values. The sections $1$ and $3$ have dimensions $L_{x} = 400a$ and $c = 50a$, where $a$ is the lattice spacing (see Fig. \ref{racetrack}). The horizontal dimension of section $2$ is also $L_{x} = 400a$ and its width  can range from $b = 1a$ to $b = 9a$. Then, the total width of the sample is $L_{y} = 2c + b$. The full sample contains $L_{x}L_{y}$ spins.

\begin{figure*}
    \centering
    \subfigure[]{\includegraphics[width=7.5cm]{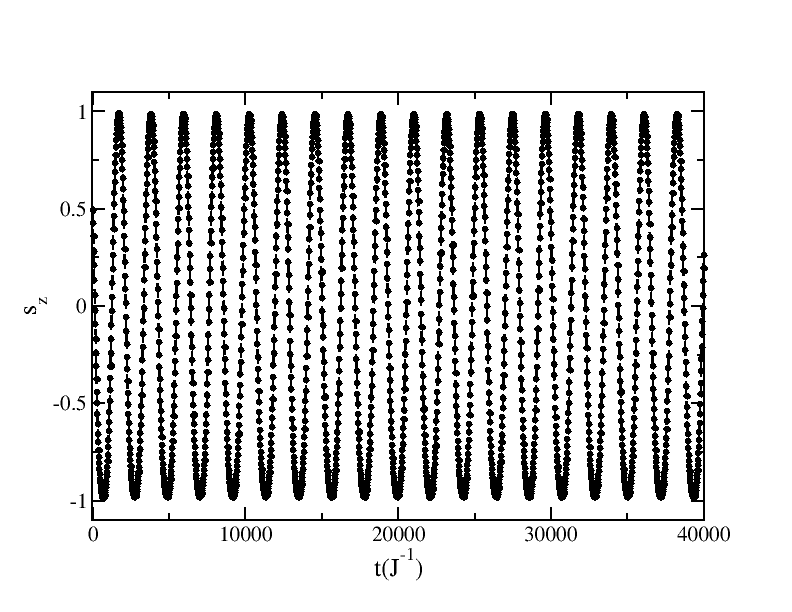}}
    \subfigure[]{\includegraphics[width=7.5cm]{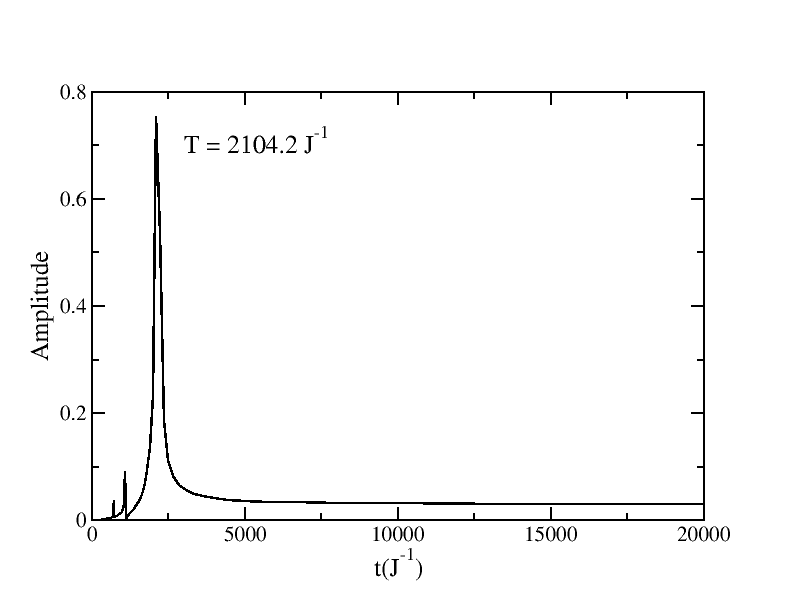}}

    \caption{(Color online) (a) Periodic oscillation of $z$-component ($S_{z}(t)$) of a spin which is located in a site near the racetrack interface. Our criterion for choosing this site is the position in which the center of the skyrmions (in a current of $SAP$ structures) is predicted to pass. Here the current has twelve $SAP$ textures. (b) Fourier transform for $S_{z}(t)$. For parameter $b=6a$, the period is $T = 2104.2 J^{-1}$. This result implies that, in this type of racetrack, the velocity of a $SAP$ current is the same as the velocity of only one $SAP$.}
     \label{oscillation}
\end{figure*}

By just using the spin dynamics technique, we obtain a system with a single skyrmion (section $1$) and a single antiskyrmion (section $3$) as shown in Fig. \ref{racetrack}. After that, we apply a Berger spin-transfer torque in the system with $\vec{\jmath}=0.10 \,\hat{x}$ along the horizontal direction of the racetrack (Fig.  \ref{current1}a). Here, we must emphasize that the current is homogeneous throughout the whole racetrack. Both the skyrmion and antiskyrmion experience Magnus force and their pathways are deflected in the direction of the sector $2$ (green zone of Fig.  \ref{current1}b). When the skyrmion and the antiskyrmion touch the central region, a bound state skyrmion-antiskyrmion emerges as depicted in Fig.\ref{current1}c. After the pair has been created (Fig. \ref{current1}.d), the current must be reduced by $10$ times for the quasi-particle to remain stable, without being disintegrated. Then, the skyrmion-antiskyrmion pair ($SAP$), under the action of a small current ($\vec{\jmath}=0.015 \,\hat{x}$), moves along a straight line trajectory in the same direction of the applied electric current (see the movie available as supplementary material online). The magnitude of $\vec{\jmath}$ performs a crucial role in the stability of SAP structure. It is necessary to have a fine control on the current intensity and, moreover, the range of its values necessary to maintain $SAP$ stable, depends on the section $2$ width $b$. For instance, for $b=6a$, we observe that for $|\vec{\jmath}|> 0.020$, the $SAP$ annihilation process takes place. On the other hand, for $|\vec{\jmath}|<0.012$ (including the case where the spin-polarized current is turned off), the SAP is dissolved, and the skyrmion ($S$) and antiskyrmion ($A$) keep their existence independently, each one in its respective region.

By studying the proposed racetrack with only one $SAP$ spin texture, it becomes possible to analyze in detail the time evolution of the $SAP$ mass center position. This can be done by observing the time evolution of the maximum value of the $z$-component of the spins in sections $1$ and $3$. This is equivalent to tracking the positions of the skyrmion and antiskyrmion over the time. Figure \ref{position} shows how the horizontal position $r_{x}$ of these structures evolves as a function of time (in units $J^{-1}$). The top of this figure exhibits  $r_{x} (t)$ for the skyrmion. As can be seen, there is a linear dependence between $r_{x}$ (black circles) and time. The red straight line represents a linear fit, $r_{x}(t) = \nu t+ \varphi$. We found $\nu \approx 0.016 aJ$ and $\varphi \approx 120.3 a$, which correspond to, respectively, the velocity in which the $SAP$ quasi-particle moves along the lattice and its initial position, measured in relation to the coordinates system presented in Fig.\ref{racetrack}. The behavior of the \textit{x}-position of the antiskyrmion is identical to the skyrmion behavior and so it is omitted.

The bottom part of Fig. \ref{position} shows the $y$-position of the skyrmion (black circles) and antiskyrmion (red squares) as a function of time. Note that the distance between them remains always constant. Thus, one should conclude that the integrants skyrmion $S$ and antiskyrmion $A$ of a $SAP$ particle moves together, as an unique particle under the effect of the applied spin-polarized current.

Figure \ref{energy} shows the temporal evolution of the energy (per spin) for racetracks with different interface (section $2$) widths $b$. In all cases, the energy increases as the skyrmion and the antiskyrmion approach to the central section. When these integrants touch section $2$, there is a rapid energy variation due to the $SAP$ formation, as illustrated in Fig.\ref{current1}.b. At this moment, we have to decrease the magnitude of the spin-polarized current density from $\vec{\jmath}=0.10 \, \hat{x}$ to $\vec{\jmath}=0.015 \,\hat{x}$, which remains constant during the rest of simulation. The wider interface sections have greater energy peaks because of the larger number of flipped spins in section $2$ at the moment the $SAP$ is created. After that, the energy decreases to a certain value and so it remains practically constant; the pair becomes stable. For the parameters used in our simulations, we do not observe the formation of $SAP$ for interface widths greater than $9a$.

We now present the results obtained by the thermodynamic approach. In this scenario, several excitations can be found in the racetrack, including cases in which the number of skyrmions and antiskyrmions is different. We do not consider these situations because the unpaired excitation eventually hinders the creation of $SAP$ structures; it may cause $SAP$ annihilation. Nevertheless, we expect that, for real lattices (much greater than the ones investigated here), this trouble will not occur due to the large density of these particles as observed in experiments. Despite this, we can generate a system with more than only one $SAP$, as shown in Fig. \ref{current}, where a sequence of $12$ $SAP$ quasi-particles reside on the racetrack. Therefore, a coherent current of $SAP$ spin textures can flow through the sample with constant velocity, as can be seen in Fig. \ref{oscillation}.a. As previously discussed, to maintain the SAP pairs stable and lively, it must be applied a spin current permanently. We have used $\vec{\jmath}=0.015\, \hat{x}$.  Indeed, Fig.\ref{oscillation} shows the temporal evolution of the $z$-component of a particular spin ($S_{z}$) of a site near section $2$. The choice of this magnetic moment must be done in such a way that it is located in a site where the skyrmion center (or antiskyrmion center) will pass during the movement of $SAP$ structures. Here, the inspected spin is localized at site $\vec{r}_{S}= \frac{L_{x}}{2} \,\hat{x} + \left(c-10a\right) \,\hat{y}$. Thus, as the $SAP$ current passes through the lattice, the observed $S_{z}$ continuously goes from $+1$ (when it is the central spin of a skyrmion), to $-1$, when it is in between two consecutive $SAP$ textures. Thereby, each amplitude of Fig. \ref{oscillation}.a indicates that a $SAP$ has passed in the site of the selected spin. After twelve peaks, all $SAP$ quasi-particles have passed through this site, and, as we use $PBC$ along the horizontal direction (\textit{x}-axis), the first $SAP$ restarts its trajectory. Figure \ref{oscillation}.b shows the Fourier transform of $S_{z}$ data. It presents a peak at a well defined period $T=2104.2\, J^{-1}$, which is the period of oscillation of the inspected $S_{z}$. Therefore, the $SAP$ velocity, $V_{SAP}$ can be estimated by:
\begin{equation} \label{velocity}
    V_{SAP} = \frac{L_{x}}{N_{SAP}T} = \frac{400 a}{12 (2104.2 J^{-1})} \approx 0.016 aJ,
\end{equation}
where $N_{SAP}=12$ is the number of $SAP$ excitations found on the racetrack considered in Fig. \ref{current}. Note that this velocity is the same as the one obtained from the single $SAP$ case. This indicates that the motion of SAP depends only of the applied electric current.

It is also possible to obtain the mean diameter of a skyrmion (or an antiskyrmion) belonging to a $SAP$ quasi-particle. Once the $SAP$ velocity is given by Eq.\ref{velocity}, the average skyrmion diameter can be obtained by the product of $V_{SAP}$ with the time interval $\Delta T_{1}=920.9\, J^{-1}$ for which $S_{z}$, in Fig. \ref{oscillation}, assumes positive values. So, $D_{sky}=V_{SAP} \Delta T_{1} \approx 15 a$. Similar results can be found analysing the antiskyrmion. The mean separation between the centers of two consecutive $SAP$ structures is given by $d_{SAP}=V_{SAP} T = \frac{L_{x}}{N_{SAP}}\approx 33 a$.

In summary, we have suggested a picture in which a stabilized skyrmion-antiskyrmion pair ($SAP$) can move in a straight line along a racetrack composed of three different ferromagnetic materials. Therefore, after becoming a pair, the mechanism with the $SAP$ structure proposed here indirectly overcomes an intrinsic difficulty usually found in ferromagnetic systems, i.e., the skyrmion Hall effect, which is a type of trouble to manipulate skyrmions for use in technologies. We have also shown that this engineered compound supports even a current of $SAP$ objects. The velocity of only one $SAP$ and the velocity of a group of $SAP$ structures are the same and depends only on the applied external current. We hope that this apparatus could be useful to suppress some part of the large demand of skyrmion manipulations in nanodevices, mainly for applications in spintronic/skyrmionic devices in which the $SAP$ textures would be the carrier information, yielding to a kind of ``saptronic'' technology.

\section{Acknowledgements}
The authors thank the Brazilian agencies CNPq, FAPEMIG
and CAPES(Finance Code 001) for financial support.

\end{document}